\documentclass{article}
\usepackage{amsmath,graphicx}
\usepackage{amsfonts}
\usepackage{multirow}
\usepackage{amsmath}
\usepackage{float}
\usepackage{verbatim}
\usepackage{hyperref}

\usepackage[preprint]{spconf}
\copyrightnotice{\copyright IEEE 2021}
\toappear{Submitted to {\it ASRU 2021, December 13-17, 2021, Cartagena, Colombia}}

\title{EFFICIENT CONFORMER: PROGRESSIVE DOWNSAMPLING AND GROUPED ATTENTION FOR AUTOMATIC SPEECH RECOGNITION}
\name{Maxime Burchi\sthanks{Work done during an internship at Orange Labs.}, Valentin Vielzeuf}
\address{Orange Labs, Cesson-S\'evign\'e, France \\ maxime.burchi@gmail.com, valentin.vielzeuf@orange.com}

\begin{document}
%
\maketitle
\begin{abstract}

The recently proposed Conformer architecture has shown state-of-the-art performances in Automatic Speech Recognition by combining convolution with attention to model both local and global dependencies. In this paper, we study how to reduce the Conformer architecture complexity with a limited computing budget, leading to a more efficient architecture design that we call Efficient Conformer. We introduce progressive downsampling to the Conformer encoder and propose a novel attention mechanism named grouped attention, allowing us to reduce attention complexity from $O(n^{2}d)$ to $O(n^{2}d / g)$ for sequence length $n$, hidden dimension $d$ and group size parameter $g$. We also experiment the use of strided multi-head self-attention as a global downsampling operation. Our experiments are performed on the LibriSpeech dataset with CTC and RNN-Transducer losses. We show that within the same computing budget, the proposed architecture achieves better performances with faster training and decoding compared to the Conformer. Our 13M parameters CTC model achieves competitive WERs of 3.6\%/9.0\% without using a language model and 2.7\%/6.7\% with an external n-gram language model on the test-clean/test-other sets while being 29\%\footnote{Inference time on a single Intel Core i9-9940X 3.3GHz CPU thread.} faster than our CTC Conformer baseline at inference and 36\% faster to train.\footnote{Code is available at \href{https://github.com/burchim/EfficientConformer}{https://github.com/burchim/EfficientConformer}.}

\end{abstract}

\section{Introduction}
\label{sec:introduction}

\begin{figure}[tb]
        \centering
        \includegraphics[width=0.75\linewidth]{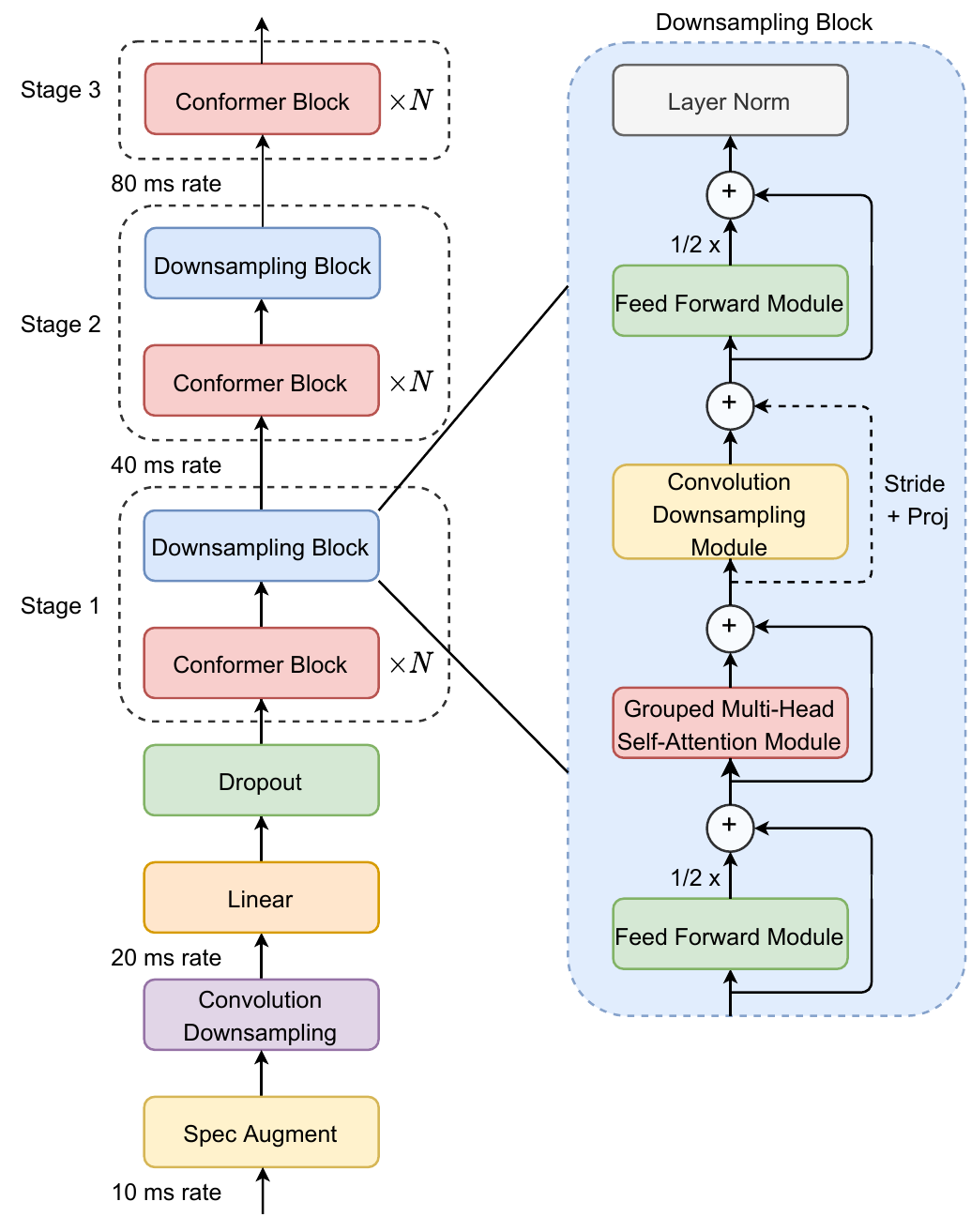}
        \caption{\textbf{Efficient Conformer encoder model architecture.} The Efficient Conformer encoder is composed of three stages where each stage comprises a number of Conformer blocks using grouped attention. Encoded sequence is progressively downsampled and projected to wider feature dimensions.}
    \label{fig:EffConf}
\end{figure}
End-to-end automatic speech recognition (ASR) has become the standard of state-of-the-art approaches. Indeed the availability of large scale hand-labeled datasets and sufficient computing resources made it possible to train powerful deep neural networks for ASR, reaching very low Word Error Rate (WER) on academic benchmarks.
Yet even if these new approaches are breaking the state-of-the-art, one major pitfall for using them in real-world is the resource cost. To achieve high performance, the training budget is often very large, implying to use a sizeable number of GPUs~\cite{kriman2020quartznet,majumdar2021citrinet}. And when the model has been successfully trained, the inference time may also become prohibitive for some specific usage. For instance, available devices often do not come with a GPU and an ideal usage in a production environment would be to be able to compute the inference on a single basic CPU. 

The integration of neural networks as a production-ready technology has been broadly explored in many fields such as vision~\cite{sandler2018mobilenetv2} and different approaches have been proposed to address these problems. They may be gathered into several broad categories~\cite{neill2020overview}, such as weights sharing~\cite{dabre2019recurrent}, pruning~\cite{cantu2003pruning}, quantization~\cite{dally2015high}, knowledge distillation~\cite{bucilua2006model}, low-rank decomposition~\cite{xue2013restructuring} and efficient architecture design~\cite{tan2019efficientnet}. Each of these methods (separately or together) may help to reduce the model complexity. In this paper we choose to focus on the design of an efficient architecture to address the ASR problem. 

Different types of architectures have been used for ASR, such as RNN~\cite{graves2013speech, hannun2014deep, chan2015listen, rao2017exploring, he2019streaming}, CNN~\cite{collobert2016wav2letter, li2019jasper, kriman2020quartznet, han2020contextnet, majumdar2021citrinet} and transformers~\cite{dong2018speech, karita2019comparative, zhang2020transformer, guo2021recent}. More recently architectures modelling both local and global dependencies have been introduced. For instance,~\cite{han2020contextnet, majumdar2021citrinet} enhance the global context of CNNs with the squeeze-and-excitation (SE) mechanism~\cite{hu2018squeeze}, while~\cite{gulati2020conformer} augment the transformer network with convolution to model both local and global dependencies with convolution and attention achieving state-of-the-art results. 
We aim at reducing this Conformer complexity while coping with some strict constraints: limiting the training resource budget to a maximum of 4 Nvidia RTX 2080 Ti GPUs without harming the recognition performance.  
Recent works done in ASR to reduce the computation cost of CNNs for faster training and inference~\cite{han2020contextnet} show the interest of applying a progressive downsampling from the bottom to the top of the model. We propose to introduce this progressive downsampling and dimension scaling to the Conformer. Following the same patterns proposed in~\cite{han2020contextnet}, we progressively reduce the length of the encoded sequence by a factor of 8. We also study the benefit of using multi-head attention as a global downsampling operation, instead of using a convolution downsampling.

Yet, one main drawback of applying a progressive subsampling approach to an attention-based architecture is the introduction of a computation asymmetry into the network. As attention complexity is quadratic in the sequence length, earlier attention layers require way more computation than latter layers and result in a time bottleneck.
The same problem is found in vision where recent works~\cite{bello2019attention, ramachandran2019stand, srinivas2021bottleneck} have proposed to replace or augment convolution with self-attention in the ResNet family backbone~\cite{he2016deep}. The adopted solution is then to restrict attention to latter layers with smallest spatial dimension and therefore hit computation and memory constraints. Yet, this may mean a performance degradation in the specific case of the Conformer.
A solution would be to build an efficient self-attention mechanism (its original form has a quadratic time complexity). Indeed, \cite{li2021efficient} shows the benefits brought by using efficient attention~\cite{shen2021efficient}, while \cite{wang2021efficient} proposes a prob-sparse mechanism to decide whether the attention operation should be computed. These approaches greatly deal with the problem of handling longer sequences, but may bring marginal improvements for small sequences~\cite{wang2021efficient} (Figure 3). Another alternative to regular attention is local attention~\cite{parmar2018image, ramachandran2019stand}, which is inspired by CNNs and restricts the positions in the attended positions to a local neighbourhood around the query position.
In this work, we take inspiration from all these approaches and propose a sequence-length agnostic attention mechanism that we call grouped attention. Grouped attention reduces attention complexity from $O(n^{2}\cdot d)$ to $O(n^{2}\cdot d / g)$ by grouping neighbouring time elements of the sequence along the feature dimension before applying scaled dot-product attention.
Therefore, this paper proposes an efficient Conformer which combines both Progressive Subsampling and Grouped Attention. We apply a stronger grouped multi-head self-attention to early attention layers in the encoder first stage, where the sequence is the longest and therefore the complexity the highest. We show that it allows to greatly reduce the computation asymmetry and thus the computation time.

Finally, the original Conformer has initially been trained with a RNN-T criteria, while works using the ESPnet toolkit \cite{li2021efficient,shen2021efficient} propose a training based on a combined CTC and Attention loss. 
Recent works have shown that it is also possible for fully convolutional models to reach great performance using the single CTC loss~\cite{majumdar2021citrinet} and thus to gain an important decoding time. We propose to better investigate these benefits in a resource constrained environment, comparing our encoder trained with CTC and with RNN-T.

This work brings four main contributions:
\textbf{(a)} the introduction of a \textbf{Progressive Downsampling} to the Conformer encoder leading to a more efficient architecture achieving better recognition performances with fewer multiply-adds,
\textbf{(b)} a novel attention mechanism that we call \textbf{Grouped Attention}, allowing us to further reduce training and decoding time of our Efficient Conformer model while maintaining similar recognition performances, \textbf{(c)} a \textbf{comparative study} of the benefits brought by training the Conformer with the \textbf{original RNN-T approach versus the CTC one}, with respect to a restricted training budget setting and \textbf{(d)} a small efficient Conformer with \textbf{competitive recognition performance}.

\begin{figure*}[ht]
    \begin{minipage}[]{1.0\linewidth}
        \centering
        \centerline{\includegraphics[width=\linewidth]{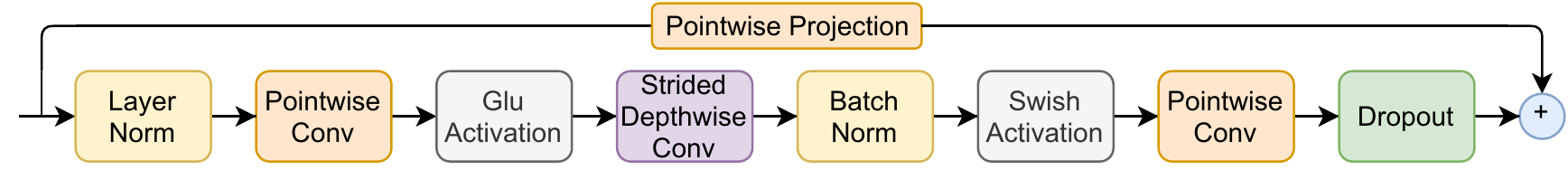}}
        \caption{\textbf{Convolution downsampling module.} Sequence downsampling is performed using a strided depthwise convolution. A pointwise convolution projects the number of channels using an expansion factor of $2\times d_{out}/d_{in}$ with a gated linear unit~\cite{dauphin2017language}.}
        \label{fig:ConvolutionDownsamplingModule}
    \end{minipage}
\end{figure*}

\section{Methods}
\label{sec:methods}
We propose two main strategies to reduce the Conformer complexity. Our first strategy is the introduction of progressive downsampling to the Conformer architecture, allowing us to reach better recognition performances and faster decoding. The second strategy aims at increasing the efficiency of earlier self-attention layers using grouped and local attention to balance model overall complexity without hurting accuracy. We experiment with CTC~\cite{graves2006connectionist} and RNN-T~\cite{graves2012sequence} losses, comparing the impact of the proposed set on methods for both criteria.

\subsection{RNN-T and CTC Criteria}
RNN-T extends CTC by defining a distribution over output sequences of all lengths, and by jointly modelling both input-output and output-output dependencies. The audio encoder (or transcription network) is combined with a label decoder (or prediction network) and a joint network~\cite{graves2013speech}. The joint network combines the audio encoder and label decoder outputs using a feed forward neural network with a softmax output layer over the vocabulary size. For CTC, the encoder is augmented with a final softmax layer that directly converts the encoder outputs to probabilities.

\subsection{Progressive Downsampling}
\label{sec:downsampling}
Inspired by recent works done in ASR to reduce the computation cost of CNNs for faster training and inference with progressive downsampling~\cite{han2020contextnet, majumdar2021citrinet}, we experiment introducing progressive downsampling to the Conformer encoder. Our Efficient Conformer encoder, illustrated in Figure~\ref{fig:EffConf}, first downsamples audio features with a $3 \times 3$ convolution stem with stride 2. The resulting features are fed to three encoder stages where each stage comprises a number of conformer blocks~\cite{gulati2020conformer} of same feature dimension. A conformer block is composed of a multi-head self-attention module and a convolution module sandwiched between two feed-forward networks. Each block is followed by a post layer normalization. Sequence downsampling is performed in the last block of first and second encoder stages. We replace the original convolution module with a convolution downsampling module illustrated in Figure~\ref{fig:ConvolutionDownsamplingModule}. We also experiment with attention downsampling using a strided attention in the multi-head self-attention module, as shown in Figure~\ref{fig:AttentionDownsamplingModule}. This results in a 8$\times$ progressive downsampling performed along the time dimension. The encoded sequence is progressively projected to wider feature dimension such that the complexity of hidden layers stay the same for each encoder stage. This is achieved in the convolution module of every downsampling block.

\subsection{Towards an efficient Self-Attention}
\textbf{Relative Multi-Head Self-Attention}
Self-attention is used to introduce global dependencies into the network by computing dot-products between each element of the hidden sequence. In the case of multi-head self-attention (MHSA)~\cite{vaswani2017attention}, a scaled dot-product attention is performed individually for a number of heads $H$ to a hidden sequence $X\in \mathbb{R}^{n \times d}$ as:
\begin{align}
MHSA(X) = Concat\left(O_{1}, ..., O_{H}\right)W^{O},\\
\text{where}\ O_{h} = softmax\left(\frac{Q_{h}K_{h}^{T}}{\sqrt{d_{h}}}\right)V_{h}
\end{align}
Where $Q_{h}=XW^{Q}_{h}$, $K_{h}=XW^{K}_{h}$ and $V_{h}=XW^{V}_{h}$ are query, key and value linear projections with parameter matrices $W^{Q}_{h}$, $W^{K}_{h}$, $W^{V}_{h}\in \mathbb{R}^{d \times d_{h}}$ and $W^{O}\in \mathbb{R}^{d \times d}$ is the output linear projection matrix. As \cite{gulati2020conformer}, we use multi-head self-attention with relative sinusoidal positional encodings, allowing the model to generalize better on different input lengths. We adapt the original relative positional encodings from Transformer-XL~\cite{dai2019transformer} to full context using a sinusoidal matrix $R \in \mathbb{R}^{(2n_{max}-1) \times d}$ with positions ranging from $-(n_{max}-1)$ to $(n_{max}-1)$. The output of a head $h$ becomes:
\begin{align}
O_{h} = softmax\left(\frac{Q_{h}K_{h}^{T}+S^{rel}_{h}}{\sqrt{d_{h}}}\right)V_{h}
\end{align}
Where $S^{rel} \in \mathbb{R}^{n \times n}$ is a relative position score matrix that satisfy $S^{rel}[i, j]=Q_{i}E_{j-i}^{T}$ with relative position embedding $E=RW^{E}$. This condition is achieved by reindexing $QE^{T}$, moving the relative logits to their correct positions. The memory efficient relative to absolute position indexing algorithm for unmasked sequences is described in \cite{bello2019attention} (Appendix A.3).

\begin{figure}[ht]
    \begin{minipage}[b]{1.0\linewidth}
        \centering
        \centerline{\includegraphics[width=\linewidth]{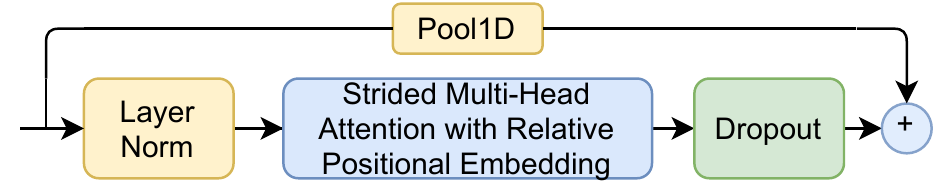}}
        \caption{\textbf{Multi-head self-attention downsampling module.} Attention downsampling is performed using a strided attention with relative position encodings and a pooling residual.}
        \label{fig:AttentionDownsamplingModule}
    \end{minipage}
    \begin{minipage}[b]{1.0\linewidth}
        \centering
        \centerline{\includegraphics[width=0.99\linewidth]{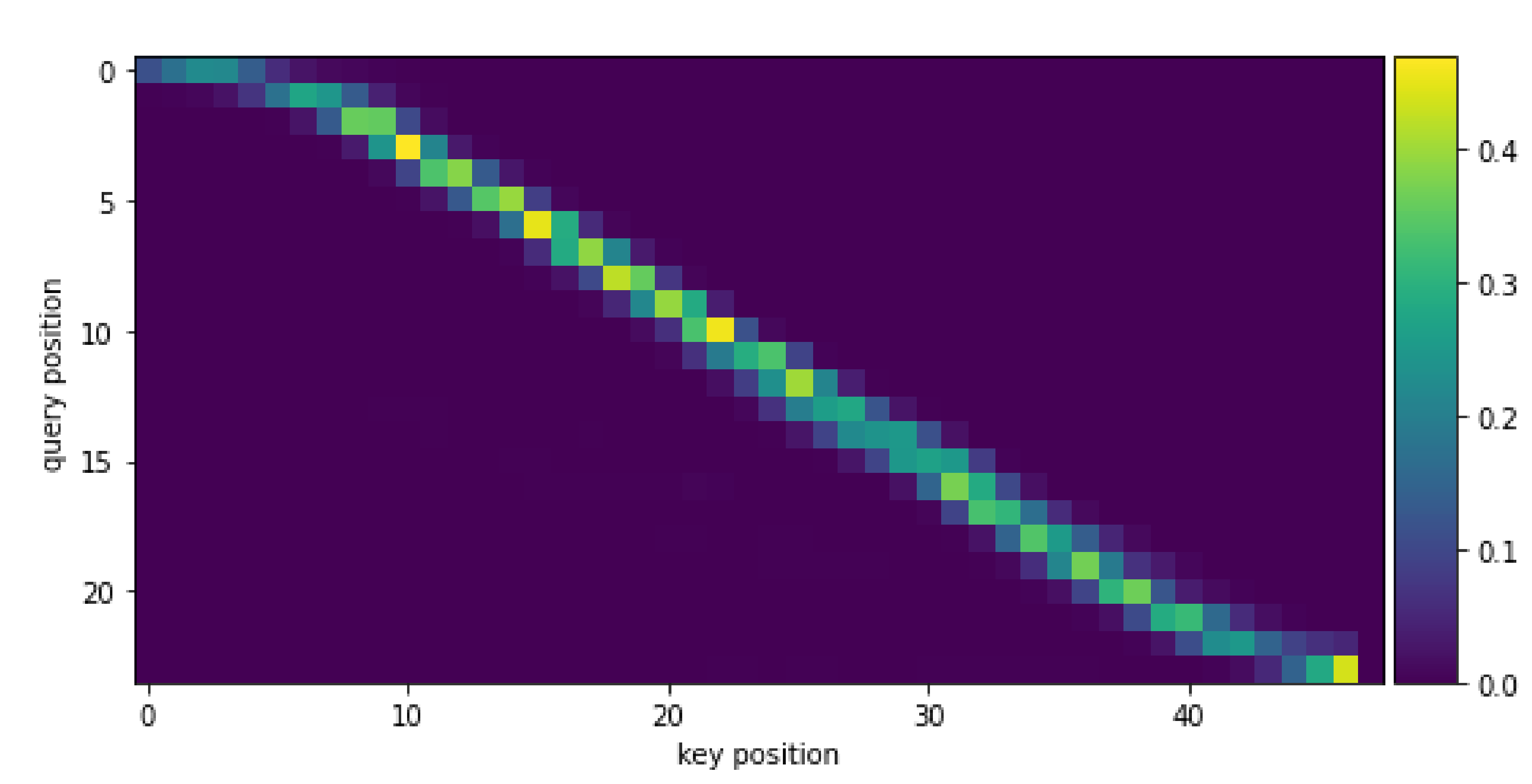}}
        \caption{Strided attention head capturing local information on the diagonal to perform downsampling. We also observed attention heads specialized for long-term relationships.}
        \label{fig:AttSubMap}
    \end{minipage}
\end{figure}

\begin{figure*}[tb]
    \begin{minipage}[]{0.99\linewidth}
        \centering
        \centerline{\includegraphics[width=\linewidth]{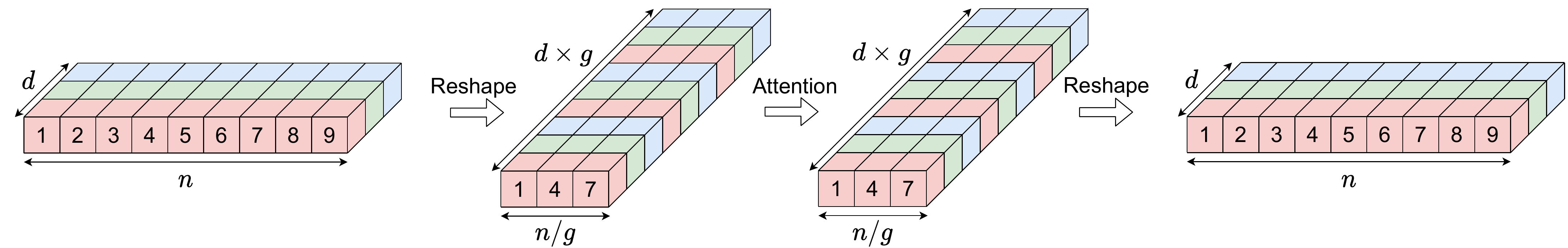}}
        \caption{\textbf{Grouped Multi-Head Attention.} Queries, keys, values and position embeddings are reshaped by grouping nearby elements along the feature dimension, reducing attention complexity from $O(n^{2}d)$ to $O(n^{2}d / g)$ where $g$ defines the number of time elements per group. Grouped attention is equivalent to regular attention when $g=1$.}
        \label{fig:GroupedAtt}
    \end{minipage}
\end{figure*}

\textbf{Strided Multi-Head Self-Attention}
Downsampling is generally performed using strided convolution or pooling operations. These operations performs local downsampling by processing nearby elements of a hidden sequence. In this work, we experiment the use of MHSA as a global downsampling operation. This is achieved by striding the attention query along the temporal dimension resulting in subsampled query $Q_{sub}\in \mathbb{R}^{n/2 \times d}$. This results in strided attention maps, as shown in Figure~\ref{fig:AttSubMap}, where subsampled query positions can attend to the entire sequence context to perform downsampling. Strided MHSA has a $O(n^{2}d/s)$ complexity where $s$ defines the stride applied to the query projection. Progressive attention downsampling is performed by replacing regular MHSA layers by strided MHSA layers in each downsampling block. Figure~\ref{fig:AttentionDownsamplingModule} illustrates the multi-head self-attention downsamping module.

\textbf{Grouped Multi-Head Self-Attention}
While a similar complexity per hidden layer can be obtained for different encoder stages by varying blocks feature dimension for each encoder stage, attention complexity is quadratic in the sequence length which introduces computation asymmetry into the network where earlier attention layers requires way more multiply-adds than latter layers. We propose to solve this problem by defining a novel attention mechanism, which we call grouped attention (Figure \ref{fig:GroupedAtt}). Grouped attention reduce attention complexity from $O(n^{2}\cdot d)$ to $O(n^{2}\cdot d / g)$ by grouping nearby time elements along the feature dimension before applying scaled dot-product attention. Attention queries, keys, values and relative positional embedding are reshaped from $Q$, $K$, $V\in \mathbb{R}^{n\times d}$ and $E\in \mathbb{R}^{(2n-g)\times d}$ to $Q^{grp}$, $K^{grp}$, $V^{grp}\in \mathbb{R}^{n'\times d'}$ and $E^{grp}\in \mathbb{R}^{(2n'-1)\times d'}$ where $n'=n/g$ and $d'=d\times g$. The output of a head $h$ becomes:
\begin{align}
O^{grp}_{h} = softmax\left(\frac{Q^{grp}_{h}K^{grpT}_{h}+S^{rel}_{h}}{\sqrt{d_{h}'}}\right)V^{grp}_{h}
\end{align}
And concatenated grouped attention output $O^{grp} \in \mathbb{R}^{n'\times d'}$ is reshaped to $O \in \mathbb{R}^{n\times d}$ before the output projection layer. Grouped multi-head self-attention is motivated by the fact that nearby element are supposed to encode similar features and therefore a low resolution attention pattern could be applied to approximate regular dense attention. We apply grouped multi-head self-attention starting from earlier attention layers in the first stage where encoded sequence is the longest before experimenting with second and third stages. 

\textbf{Local Multi-Head Self-Attention}
Introduced in~\cite{parmar2018image, ramachandran2019stand}, local attention restricts the attended positions to a local neighborhood around the query position. This is achieved by defining an attention window $w_{att}$ and segmenting the hidden sequence into blocks of size $w_{att}$. 
Regular MHSA is then performed in parallel for each block where all queries attends to the same content matrix comprised of all block positions. 

\section{Experiments}
\label{sec:Experiments}

\subsection{Data and Training Setup}
\textbf{Data} We train and evaluate our models on the LibriSpeech \cite{panayotov2015librispeech} dataset. LibriSpeech is a corpus of approximately 1000 hours of 16kHz read English speech with corresponding text transcripts. An additional 800 millions token text-only corpus is provided for language model (LM) training. We use input spectrograms of 80-dimensional mel-scale log filter banks computed over windows of 20ms strided by 10ms. SpecAugment~\cite{park2020specaugment} is applied during training to prevent overfitting with two frequency masks with mask size parameter $F=27$ and ten time masks with adaptive size $p_{S} = 0.05$. We only use five time masks for CTC experiments as our models failed to converge with ten masks. 

\textbf{Training Setup}
We experiment with RNN-Transducer and CTC  models of 10M and 13M parameters respectively. Table~\ref{table:hp} describes our models hyper-parameters. Transducer models use a single LSTM layer decoder. The decoder and joint network dimensions are set to 320 in every experiments. A byte-pair encoding tokenizer is built from LibriSpeech transcripts using sentencepiece~\cite{kudo2018sentencepiece}. Following previous works~\cite{gulati2020conformer, majumdar2021citrinet}, we use a 1k subwords lexicon size for transducer models and 256 for CTC. All models were implemented from scratch in PyTorch~\cite{pytorch}.

We train CTC models for 450 epochs with a global batch size of 256 on 4 GPUs, using a batch size of 32 per GPU with 2 accumulated steps. Transducer models are trained for 250 epochs using batch sizes of 16 per GPU and 4 accumulated steps. We use the Adam optimizer~\cite{kingma2014adam} with $\beta_1$ = 0.9, $\beta_2$ = 0.98, $\epsilon$ = $10^{-9}$ and a transformer learning rate schedule~\cite{vaswani2017attention} with 10k warmup-steps and peak learning rate $0.02/\sqrt{d_{enc}}$ and $0.05/\sqrt{d_{enc}}$ for CTC and RNN-T respectively, where $d_{enc}$ is the encoder output dimension. Gaussian weight noise~\cite{jim1996analysis} ($\mu$ = 0, $\sigma$ = 0.075) is added to the transducer decoder during training for regularization starting at 20k steps, re-sampling the noise at every training step. We also add a L2 regularization with a $1e^{-6}$ weight to all the trainable weights of the model. 

We train a 6-gram external language model~\cite{heafield2011kenlm} on the LibriSpeech LM corpus for re-scoring during beam search.

\begin{table}[htb]
\centering
\caption{Conformer and Efficient Conformer (Eff Conf) models hyper-parameters for CTC and RNN-T experiments.}
\scriptsize
\begin{tabular}{c|cccc}
\hline
\multirow{2}{*}{Model} & \multicolumn{1}{c}{\multirow{2}{*}{\begin{tabular}[c]{@{}c@{}}Conformer\\ Transducer\end{tabular}}} & \multicolumn{1}{c}{\multirow{2}{*}{\begin{tabular}[c]{@{}c@{}}Eff Conf\\ Transducer\end{tabular}}} & \multicolumn{1}{c}{\multirow{2}{*}{\begin{tabular}[c]{@{}c@{}}Conformer\\ CTC\end{tabular}}} & \multicolumn{1}{c}{\multirow{2}{*}{\begin{tabular}[c]{@{}c@{}}Eff Conf\\ CTC\end{tabular}}} \\ \\ \hline\hline
Num Params (M) & 10.3          & 10.8            & 13.0       & 13.2        \\
Encoder Blocks   & 16            & 5,5,5           & 16         & 5,5,5       \\
Encoder Dims  & 144           & 100,140,200     & 176        & 120,168,240 \\
Attention Heads  & 4             & 4,4,4           & 4          & 4,4,4       \\
Conv Kernel Size     & 31            & 15,15,15        & 31         & 15,15,15 \\
Att Group Size     & -            & 3,1,1        & -         & 3,1,1 \\\hline
\end{tabular}
\label{table:hp}
\end{table}

\subsection{Results on LibriSpeech}
Table~\ref{table:results} compares the Word Error Rates (WER) of our experiments with state-of-the-art CTC (QuartzNet, CitriNet) and RNN-T (Conformer Transducer, ContextNet) models on the LibriSpeech test-clean and test-other sets. Our Efficient Conformer CTC model achieves competitive results of 3.57/8.99 without a language model for only 13M parameters. It even outperforms the 21M parameter Citrinet-384 using an external 6-gram language model during beam search, achieving WERs of 2.72/6.66. Moreover, we were able to recover similar results compared to the non-grouped version with 35\% faster training using grouped attention with parameter $g=3$ in the first stage. Our Efficient Conformer Transducer model achieves satisfying results but still lack behind the original work that was trained with larger batches and more resources. We found RNN-T models to converge faster with fewer epochs than CTC models, achieving lower greedy WER. However, using an external language model during beam search allows CTC models to bridge the gap in WER with RNN-T, which is in line with what was observed in \cite{majumdar2021citrinet}.
\begin{table}[ht]
\centering
\scriptsize
\caption{Comparison of LibriSpeech WER(\%) with recent published RNN-T and CTC models.}
\hfill \break
\begin{tabular}{cccccc}
\hline
\multicolumn{1}{c}{\multirow{2}{*}{\begin{tabular}[c]{@{}c@{}}\textbf{Model} \\ \textbf{Architecture}\end{tabular}}} & \multicolumn{1}{c}{\multirow{2}{*}{\begin{tabular}[c]{@{}c@{}}\textbf{Model} \\ \textbf{Type}\end{tabular}}} & \multirow{2}{*}{\textbf{LM}} & \multicolumn{2}{c}{\textbf{test WER}} & \multirow{2}{*}{\begin{tabular}[c]{@{}c@{}}\textbf{Params}\\ \textbf{(M)}\end{tabular}} \\ \cline{4-5}
\multicolumn{1}{c}{}                       &                           &  & \textbf{clean}       & \textbf{other}   &       \\ \hline
QuartzNet-15x5\cite{kriman2020quartznet}  & CTC           & -         & 3.90  & 11.28 &       \\
                &               & 6-gram    & 2.96  & 8.07  & 19    \\
                &               & Trans-XL  & 2.69  & 7.25  &       \\ \hline
Citrinet-256\cite{majumdar2021citrinet}    & CTC           & -         & 3.78  & 9.60   &       \\
                &               & 6-gram    & 3.65  & 8.06  & 9.8   \\
                &               & Trans-XL  & 2.75  & 6.87  &       \\ \hline
Citrinet-384\cite{majumdar2021citrinet}    & CTC           & -         & 3.20  & 7.90  &       \\
                &               & 6-gram    & 2.94  & 6.71  & 21.0  \\
                &               & Trans-XL  & 2.52  & 5.95  &       \\ \hline
ContextNet(S)\cite{han2020contextnet}   & RNN-T    & -         & 2.90  & 7.00  & 10.8  \\
                &               & RNN       & 2.3   & 5.5   &       \\ \hline
Conformer(S)\cite{gulati2020conformer}    & RNN-T    & -         & 2.70  & 6.30  & 10.3  \\
                &               & RNN       & 2.1   & 5.0   &       \\ \hline\hline               
Conformer(ours) & CTC           & -         & 4.07  & 10.25 & 13.0\\
                &               & 6-gram    & 2.88  & 7.25\\    \hline
Eff Conformer   & CTC           & -         & 3.58  & 8.88  & 13.2\\ 
w/o Grouped Att                &               & 6-gram    & 2.79  & \textbf{6.65}\\\hline
Eff Conformer   & CTC           & -         & 3.57  & 8.99  & 13.2\\ 
   &               & 6-gram    & \textbf{2.72}  & 6.66\\\hline
Conformer(ours) & RNN-T    & -         & 3.31 & 8.34 & 10.3\\
                &               & 6-gram    & 3.01 & 7.58 \\    \hline
Eff Conformer   & RNN-T    & -         & 3.25 & 8.08 & 10.8\\
w/o Grouped Att                &               & 6-gram    & 2.79 & 7.03 \\\hline
Eff Conformer   & RNN-T    & -         & 3.28 & 8.03 & 10.8\\
   &           & 6-gram    & 2.83 & 7.05 \\ \hline
\end{tabular}
\label{table:results}
\end{table}

\subsection{Ablation Studies}
We propose a detailed ablation study to better understand the improvements (in terms of complexity reduction and WER) brought by the different methods composing the Efficient Conformer. We report the number of operations measured by multiply-adds (MAdds) for the encoder to process a ten second audio clip. Inverse Real Time Factor (Inv RTF) is measured on the LibriSpeech dev-clean set by decoding with a batch size 1 on a single Intel Core i9-9940X 3.3GHz CPU thread. We also report experiments training time on 4 Nvidia RTX 2080 Ti GPUs.

\textbf{Progressive Downsampling}
We first study the impact of using progressive downsampling with regular MHSA in every stage. Although having a significant computation overhead in earlier layers applying MHSA on long sequences, a progressively downsampled architecture achieves better accuracy with fewer multiply-adds as well as shorter training and decoding time for both CTC and RNN-T experiments. We observe an improvement in WER especially on the dev-other set, as show in Table \ref{table:downsampling}. These benefits extends to the self-attention models what has already been observed for fully convolutional models in ASR~\cite{han2020contextnet}.
\begin{table}[ht]
\centering
\scriptsize
\caption{Ablation study on progressive downsampling}
\hfill \break
\begin{tabular}{ccccccc}
\hline
\multicolumn{1}{c}{\multirow{2}{*}{\begin{tabular}[c]{@{}c@{}}\textbf{Model} \\ \textbf{Architecture}\end{tabular}}} & \multicolumn{1}{c}{\multirow{2}{*}{\begin{tabular}[c]{@{}c@{}}\textbf{Model} \\ \textbf{Type}\end{tabular}}} & \multirow{2}{*}{\begin{tabular}[c]{@{}c@{}}\textbf{dev}\\ \textbf{clean}\end{tabular}} & \multirow{2}{*}{\begin{tabular}[c]{@{}c@{}}\textbf{dev}\\ \textbf{other}\end{tabular}} & \multirow{2}{*}{\begin{tabular}[c]{@{}c@{}}\textbf{MAdds} \\ \textbf{(B)}\end{tabular}} & \multirow{2}{*}{\begin{tabular}[c]{@{}c@{}}\textbf{Inv} \\ \textbf{RTF}\end{tabular}} & \multirow{2}{*}{\begin{tabular}[c]{@{}c@{}}\textbf{Train} \\ \textbf{Time (h)}\end{tabular}}\\\\ \hline
Conformer & RNN-T    & 3.18  & 8.42  & 3.73  & 38.5  & 158            \\
+ Prog Down     & RNN-T    & \textbf{3.13}  & \textbf{8.05}  & \textbf{2.84}  & \textbf{43.2} &  \textbf{147}           \\\hline
Conformer & CTC           & 3.81  & 10.47 & 5.41  & 44.0  & 195             \\
+ Prog Down     & CTC           & \textbf{3.44}  & \textbf{9.12}  & \textbf{3.91}  & \textbf{48.8} & \textbf{191}              \\\hline
\end{tabular}
\label{table:downsampling}
\end{table}

\textbf{Downsampling: Convolution VS Attention}
We experiment with two downsampling methods, a local downsampling performed with strided depthwise convolution in the convolution module and a global downsampling performed by strided attention. As seen in Table~\ref{table:type}, we find attention downsampling to perform as well as convolution downsampling. Furthermore, attention downsampling slightly reduces the decoding time of our Efficient Conformer model. This show that MHSA can successfully be applied as a global downsampling operation to reduce the encoded sequence length.
\begin{table}[ht]
\centering
\scriptsize
\caption{Ablation study on downsampling method}
\hfill \break
\begin{tabular}{ccccccc}
\hline
\multicolumn{1}{c}{\multirow{2}{*}{\begin{tabular}[c]{@{}c@{}}\textbf{Downsampling} \\ \textbf{Method}\end{tabular}}} & \multicolumn{1}{c}{\multirow{2}{*}{\begin{tabular}[c]{@{}c@{}}\textbf{Model} \\ \textbf{Type}\end{tabular}}} & \multirow{2}{*}{\begin{tabular}[c]{@{}c@{}}\textbf{dev}\\ \textbf{clean}\end{tabular}} & \multirow{2}{*}{\begin{tabular}[c]{@{}c@{}}\textbf{dev}\\ \textbf{other}\end{tabular}} & \multirow{2}{*}{\begin{tabular}[c]{@{}c@{}}\textbf{MAdds} \\ \textbf{(B)}\end{tabular}} & \multirow{2}{*}{\begin{tabular}[c]{@{}c@{}}\textbf{Inv} \\ \textbf{RTF}\end{tabular}} & \multirow{2}{*}{\begin{tabular}[c]{@{}c@{}}\textbf{Train} \\ \textbf{Time (h)}\end{tabular}}\\\\ \hline
Convolution & RNN-T & 3.13  & 8.05  & 2.84  & 43.2  & 147            \\
Attention   & RNN-T & \textbf{3.09}  & \textbf{7.90}  & \textbf{2.75}  & \textbf{44.6} & \textbf{144}            \\\hline
Convolution & CTC   & 3.44 & \textbf{9.12} & 3.91  & 48.8 & 191              \\
Attention   & CTC   & \textbf{3.41} & 9.23 & \textbf{3.79}  & \textbf{49.7}   & \textbf{184}             \\\hline
\end{tabular}
\label{table:type}
\end{table}

\textbf{Attention Group Size}
\label{sec:AblationGroupSize}
To study the effect of grouped attention on model complexity and recognition performance, we experiment to gradually increase attention group size in each encoder stage. The results in Table~\ref{table:group} demonstrate the effectiveness of using grouped attention in earlier attention layers to reduce model complexity and memory cost without impacting recognition performances. Introducing grouped attention in the first stage of our progressively downsampled Conformer CTC model results in a 21\% speedup in inference time with 35\% faster training. Inference time can further be reduced to 29\% speedup with 41\% faster training by introducing multi-head grouped attention in every stage but results in small performance losses. 
\begin{table}[ht]
\centering
\scriptsize
\caption{Ablation study on attention group size}
\hfill \break
\begin{tabular}{ccccccc}
\hline
\multicolumn{1}{c}{\multirow{2}{*}{\begin{tabular}[c]{@{}c@{}}\textbf{Attention} \\ \textbf{Group Sizes}\end{tabular}}} & \multicolumn{1}{c}{\multirow{2}{*}{\begin{tabular}[c]{@{}c@{}}\textbf{Model} \\ \textbf{Type}\end{tabular}}} & \multirow{2}{*}{\begin{tabular}[c]{@{}c@{}}\textbf{dev}\\ \textbf{clean}\end{tabular}} & \multirow{2}{*}{\begin{tabular}[c]{@{}c@{}}\textbf{dev}\\ \textbf{other}\end{tabular}} & \multirow{2}{*}{\begin{tabular}[c]{@{}c@{}}\textbf{MAdds} \\ \textbf{(B)}\end{tabular}} & \multirow{2}{*}{\begin{tabular}[c]{@{}c@{}}\textbf{Inv} \\ \textbf{RTF}\end{tabular}} & \multirow{2}{*}{\begin{tabular}[c]{@{}c@{}}\textbf{Train} \\ \textbf{Time (h)}\end{tabular}} \\\\ \hline
1,1,1   & RNN-T  &  3.13     &   8.05    & 2.84    & 43.2  & 147            \\
3,1,1   & RNN-T  &  2.99     &   8.27    & 2.51    & 49.1 & 103              \\\hline
1,1,1   & CTC     & 3.44  & 9.12  & 3.91    & 48.8 & 191              \\
3,1,1   & CTC    & 3.40  & 9.13  & 3.51    & 61.9 & 124               \\
5,3,1   & CTC   & 3.39  & 9.64  & 3.29    & 65.5 & 120             \\
9,5,3   & CTC  &  3.56  &   9.74    & 3.16    & 68.9 & 113              \\\hline
\end{tabular}
\label{table:group}
\end{table}

\textbf{Local Attention}
\label{sec:AblationLocalAttention}
We study the impact of local self-attention on recognition performances and inference time. Table \ref{table:local} shows the results obtained for introducing local attention in encoder stages. We find local attention to perform similarly compared to the regular multi-head attention using a local attention window $w_{att}=175$ in the first stage. However, further restricting the size of the attention window can negatively impact recognition performances. These results show the importance of using a global context for better recognition performances. It also may explain why grouped attention achieves better results. Moreover, we find local attention to be slower than grouped attention at decoding time due to the computation overhead introduced by sequence padding for partitioning sequences into non overlapping blocks of size $w_{att}$. 
\begin{table}[ht]
\centering
\scriptsize
\caption{Ablation study on local attention window}
\hfill \break
\begin{tabular}{ccccccc}
\hline
 \multicolumn{1}{c}{\multirow{2}{*}{\begin{tabular}[c]{@{}c@{}}\textbf{Attention} \\ \textbf{Window}\end{tabular}}} & \multicolumn{1}{c}{\multirow{2}{*}{\begin{tabular}[c]{@{}c@{}}\textbf{Model} \\ \textbf{Type}\end{tabular}}} & \multirow{2}{*}{\begin{tabular}[c]{@{}c@{}}\textbf{dev}\\ \textbf{clean}\end{tabular}} & \multirow{2}{*}{\begin{tabular}[c]{@{}c@{}}\textbf{dev}\\ \textbf{other}\end{tabular}} & \multirow{2}{*}{\begin{tabular}[c]{@{}c@{}}\textbf{MAdds} \\ \textbf{(B)}\end{tabular}} & \multirow{2}{*}{\begin{tabular}[c]{@{}c@{}}\textbf{Inv} \\ \textbf{RTF}\end{tabular}} & \multirow{2}{*}{\begin{tabular}[c]{@{}c@{}}\textbf{Train} \\ \textbf{Time (h)}\end{tabular}}\\\\ \hline
-,-,- & RNN-T & 3.13  & 8.05 & 2.84 & 43.2 & 147 \\
175,-,- & RNN-T & 3.12 & 8.19  & 2.49  & 48.5 & 99 \\\hline
 -,-,-      & CTC   & 3.44  & 9.12  & 3.91  & 48.8 & 191\\
175,-,-     & CTC   & 3.46  & 9.49  & 3.49  & 57.0 & 128\\
130,130,-   & CTC   & 3.65  & 10.10  & 3.29  &  58.9 & 119\\
100,100,100 & CTC   & 3.96  & 10.78  & 3.21  &  60.2 & 107\\\hline
\end{tabular}
\label{table:local}
\end{table}

\subsection{Models Complexity on Long Sequences}
\begin{figure}[ht]
    \begin{minipage}[]{0.99\linewidth}
        \centering
        \centerline{\includegraphics[width=\linewidth]{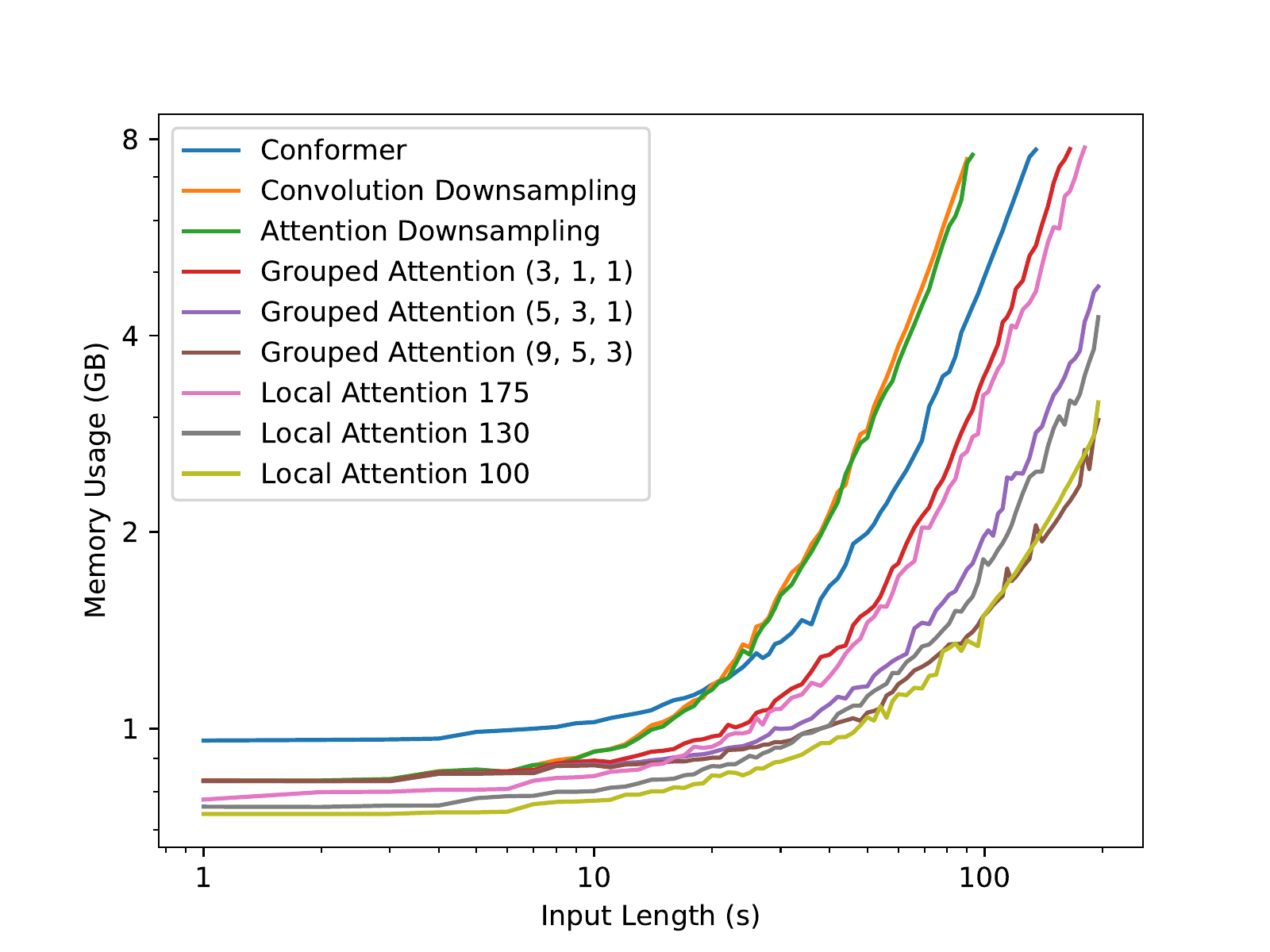}}
        \caption{CTC models memory usage for processing long sequences measured on a Nvidia GTX 1080 GPU.}
        \label{fig:memory}
    \end{minipage}
\end{figure}

Figure \ref{fig:memory} shows the impact of using progressive downsampling and attention variants on memory usage for different sequence lengths. It confirms that a progressively downsampled Conformer architecture can effectively reduce overall memory consumption when sequence length do not grow too large. However, very long sequences can result in higher memory usage due the quadratic cost of applying MHSA in earlier layers. This can be solved using efficient attention variants like local or grouped attention in earlier layers. Grouped multi-head attention can significantly reduce memory consumption for long sequences by being applied in every stages, outperforming our Conformer model using regular MHSA.

\section{Conclusion}
\label{sec:Conclusion}
In this paper, we proposed a set of methods to reduce the Conformer complexity, leading to a more efficient architecture design, the Efficient Conformer. We showed that progressive downsampling could effectively be introduced to convolution-augmented transformer networks and results in better recognition performances and faster decoding.
Then we solved the computation asymmetry caused by attention in earlier layers using a novel attention mechanism named grouped attention.
Moreover, we successfully applied strided multi-head self-attention as a global downsampling operation, achieving similar accuracy while being faster compared to convolution downsampling. 
Finally, we demonstrated the effectiveness of our methods by conducting detailed ablations studies on the LibriSpeech dataset. Our 13M parameters Efficient Conformer CTC model achieves competitive performance of 2.7\%/6.7\% for test-clean/test-other when trained on a limited computing budget of 4 GPUs while being 29\% faster than our CTC Conformer baseline at inference and 36\% faster to train.

In the future, we would like to explore other forms of attentions such as efficient attention. We also plan to apply complementary complexity reduction techniques like weights pruning and quantization to further reduce inference time.

\bibliographystyle{IEEEbib}
\bibliography{refs}

\begin{thebibliography}{10}

\bibitem{kriman2020quartznet}
Samuel Kriman, Stanislav Beliaev, Boris Ginsburg, Jocelyn Huang, Oleksii
  Kuchaiev, Vitaly Lavrukhin, Ryan Leary, Jason Li, and Yang Zhang,
\newblock ``Quartznet: Deep automatic speech recognition with 1d time-channel
  separable convolutions,''
\newblock in {\em ICASSP}, 2020, pp. 6124--6128.

\bibitem{majumdar2021citrinet}
Somshubra Majumdar, Jagadeesh Balam, Oleksii Hrinchuk, Vitaly Lavrukhin, Vahid
  Noroozi, and Boris Ginsburg,
\newblock ``Citrinet: Closing the gap between non-autoregressive and
  autoregressive end-to-end models for automatic speech recognition,''
\newblock {\em arXiv preprint arXiv:2104.01721}, 2021.

\bibitem{sandler2018mobilenetv2}
Mark Sandler, Andrew Howard, Menglong Zhu, Andrey Zhmoginov, and Liang-Chieh
  Chen,
\newblock ``Mobilenetv2: Inverted residuals and linear bottlenecks,''
\newblock in {\em CVPR}, 2018, pp. 4510--4520.

\bibitem{neill2020overview}
James~O' Neill,
\newblock ``An overview of neural network compression,''
\newblock {\em arXiv preprint arXiv:2006.03669}, 2020.

\bibitem{dabre2019recurrent}
Raj Dabre and Atsushi Fujita,
\newblock ``Recurrent stacking of layers for compact neural machine translation
  models,''
\newblock in {\em AAAI}, 2019, pp. 6292--6299.

\bibitem{cantu2003pruning}
Erick Cant{\'u}-Paz,
\newblock ``Pruning neural networks with distribution estimation algorithms,''
\newblock in {\em Genetic and Evolutionary Computation Conference}, 2003, pp.
  790--800.

\bibitem{dally2015high}
William Dally,
\newblock ``High-performance hardware for machine learning,''
\newblock {\em NeurIPS Tutorial}, vol. 2, 2015.

\bibitem{bucilua2006model}
Cristian Buciluǎ, Rich Caruana, and Alexandru Niculescu-Mizil,
\newblock ``Model compression,''
\newblock in {\em SIGKDD}, 2006, pp. 535--541.

\bibitem{xue2013restructuring}
Jian Xue, Jinyu Li, and Yifan Gong,
\newblock ``Restructuring of deep neural network acoustic models with singular
  value decomposition.,''
\newblock in {\em INTERSPEECH}, 2013, pp. 2365--2369.

\bibitem{tan2019efficientnet}
Mingxing Tan and Quoc Le,
\newblock ``Efficientnet: Rethinking model scaling for convolutional neural
  networks,''
\newblock in {\em ICML}, 2019, pp. 6105--6114.

\bibitem{graves2013speech}
Alex Graves, Abdel-rahman Mohamed, and Geoffrey Hinton,
\newblock ``Speech recognition with deep recurrent neural networks,''
\newblock in {\em ICASSP}, 2013, pp. 6645--6649.

\bibitem{hannun2014deep}
Awni Hannun, Carl Case, Jared Casper, Bryan Catanzaro, Greg Diamos, Erich
  Elsen, Ryan Prenger, Sanjeev Satheesh, Shubho Sengupta, Adam Coates, et~al.,
\newblock ``Deep speech: Scaling up end-to-end speech recognition,''
\newblock {\em arXiv preprint arXiv:1412.5567}, 2014.

\bibitem{chan2015listen}
William Chan, Navdeep Jaitly, Quoc~V Le, and Oriol Vinyals,
\newblock ``Listen, attend and spell,''
\newblock in {\em ICASSP}, 2016, pp. 4960--4964.

\bibitem{rao2017exploring}
Kanishka Rao, Ha{\c{s}}im Sak, and Rohit Prabhavalkar,
\newblock ``Exploring architectures, data and units for streaming end-to-end
  speech recognition with rnn-transducer,''
\newblock in {\em ASRU}, 2017, pp. 193--199.

\bibitem{he2019streaming}
Yanzhang He, Tara~N Sainath, Rohit Prabhavalkar, Ian McGraw, Raziel Alvarez,
  Ding Zhao, David Rybach, Anjuli Kannan, Yonghui Wu, Ruoming Pang, et~al.,
\newblock ``Streaming end-to-end speech recognition for mobile devices,''
\newblock in {\em ICASSP}, 2019, pp. 6381--6385.

\bibitem{collobert2016wav2letter}
Ronan Collobert, Christian Puhrsch, and Gabriel Synnaeve,
\newblock ``Wav2letter: an end-to-end convnet-based speech recognition
  system,''
\newblock {\em arXiv preprint arXiv:1609.03193}, 2016.

\bibitem{li2019jasper}
Jason Li, Vitaly Lavrukhin, Boris Ginsburg, Ryan Leary, Oleksii Kuchaiev,
  Jonathan~M Cohen, Huyen Nguyen, and Ravi~Teja Gadde,
\newblock ``Jasper: An end-to-end convolutional neural acoustic model,''
\newblock in {\em INTERSPEECH}, 2019, pp. 71--75.

\bibitem{han2020contextnet}
Wei Han, Zhengdong Zhang, Yu~Zhang, Jiahui Yu, Chung-Cheng Chiu, James Qin,
  Anmol Gulati, Ruoming Pang, and Yonghui Wu,
\newblock ``Contextnet: Improving convolutional neural networks for automatic
  speech recognition with global context,''
\newblock in {\em INTERSPEECH}, 2020, pp. 3610--3614.

\bibitem{dong2018speech}
Linhao Dong, Shuang Xu, and Bo~Xu,
\newblock ``Speech-transformer: a no-recurrence sequence-to-sequence model for
  speech recognition,''
\newblock in {\em ICASSP}, 2018, pp. 5884--5888.

\bibitem{karita2019comparative}
Shigeki Karita, Nanxin Chen, Tomoki Hayashi, Takaaki Hori, Hirofumi Inaguma,
  Ziyan Jiang, Masao Someki, Nelson Enrique~Yalta Soplin, Ryuichi Yamamoto,
  Xiaofei Wang, et~al.,
\newblock ``A comparative study on transformer vs rnn in speech applications,''
\newblock in {\em ASRU}. IEEE, 2019, pp. 449--456.

\bibitem{zhang2020transformer}
Qian Zhang, Han Lu, Hasim Sak, Anshuman Tripathi, Erik McDermott, Stephen Koo,
  and Shankar Kumar,
\newblock ``Transformer transducer: A streamable speech recognition model with
  transformer encoders and rnn-t loss,''
\newblock in {\em ICASSP}, 2020, pp. 7829--7833.

\bibitem{guo2021recent}
Pengcheng Guo, Florian Boyer, Xuankai Chang, Tomoki Hayashi, Yosuke Higuchi,
  Hirofumi Inaguma, Naoyuki Kamo, Chenda Li, Daniel Garcia-Romero, Jiatong Shi,
  et~al.,
\newblock ``Recent developments on espnet toolkit boosted by conformer,''
\newblock in {\em ICASSP}, 2021, pp. 5874--5878.

\bibitem{hu2018squeeze}
Jie Hu, Li~Shen, and Gang Sun,
\newblock ``Squeeze-and-excitation networks,''
\newblock in {\em CVPR}, 2018, pp. 7132--7141.

\bibitem{gulati2020conformer}
Anmol Gulati, James Qin, Chung-Cheng Chiu, Niki Parmar, Yu~Zhang, Jiahui Yu,
  Wei Han, Shibo Wang, Zhengdong Zhang, Yonghui Wu, et~al.,
\newblock ``Conformer: Convolution-augmented transformer for speech
  recognition,''
\newblock in {\em INTERSPEECH}, 2020.

\bibitem{bello2019attention}
Irwan Bello, Barret Zoph, Ashish Vaswani, Jonathon Shlens, and Quoc~V Le,
\newblock ``Attention augmented convolutional networks,''
\newblock in {\em ICCV}, 2019, pp. 3286--3295.

\bibitem{ramachandran2019stand}
Prajit Ramachandran, Niki Parmar, Ashish Vaswani, Irwan Bello, Anselm Levskaya,
  and Jonathon Shlens,
\newblock ``Stand-alone self-attention in vision models,''
\newblock {\em arXiv preprint arXiv:1906.05909}, 2019.

\bibitem{srinivas2021bottleneck}
Aravind Srinivas, Tsung-Yi Lin, Niki Parmar, Jonathon Shlens, Pieter Abbeel,
  and Ashish Vaswani,
\newblock ``Bottleneck transformers for visual recognition,''
\newblock in {\em CVPR}, 2021, pp. 16519--16529.

\bibitem{he2016deep}
Kaiming He, Xiangyu Zhang, Shaoqing Ren, and Jian Sun,
\newblock ``Deep residual learning for image recognition,''
\newblock in {\em CVPR}, 2016, pp. 770--778.

\bibitem{li2021efficient}
Shengqiang Li, Menglong Xu, and Xiao-Lei Zhang,
\newblock ``Efficient conformer-based speech recognition with linear
  attention,''
\newblock {\em arXiv preprint arXiv:2104.06865}, 2021.

\bibitem{shen2021efficient}
Zhuoran Shen, Mingyuan Zhang, Haiyu Zhao, Shuai Yi, and Hongsheng Li,
\newblock ``Efficient attention: Attention with linear complexities,''
\newblock in {\em WACV}, 2021, pp. 3531--3539.

\bibitem{wang2021efficient}
Xiong Wang, Sining Sun, Lei Xie, and Long Ma,
\newblock ``Efficient conformer with prob-sparse attention mechanism for
  end-to-endspeech recognition,''
\newblock {\em arXiv preprint arXiv:2106.09236}, 2021.

\bibitem{parmar2018image}
Niki Parmar, Ashish Vaswani, Jakob Uszkoreit, Lukasz Kaiser, Noam Shazeer,
  Alexander Ku, and Dustin Tran,
\newblock ``Image transformer,''
\newblock in {\em ICML}, 2018, pp. 4055--4064.

\bibitem{dauphin2017language}
Yann~N Dauphin, Angela Fan, Michael Auli, and David Grangier,
\newblock ``Language modeling with gated convolutional networks,''
\newblock in {\em ICML}, 2017, pp. 933--941.

\bibitem{graves2006connectionist}
Alex Graves, Santiago Fern{\'a}ndez, Faustino Gomez, and J{\"u}rgen
  Schmidhuber,
\newblock ``Connectionist temporal classification: labelling unsegmented
  sequence data with recurrent neural networks,''
\newblock in {\em ICML}, 2006, pp. 369--376.

\bibitem{graves2012sequence}
Alex Graves,
\newblock ``Sequence transduction with recurrent neural networks,''
\newblock {\em arXiv preprint arXiv:1211.3711}, 2012.

\bibitem{vaswani2017attention}
Ashish Vaswani, Noam Shazeer, Niki Parmar, Jakob Uszkoreit, Llion Jones,
  Aidan~N Gomez, Lukasz Kaiser, and Illia Polosukhin,
\newblock ``Attention is all you need,''
\newblock in {\em NeurIPS}, 2017, pp. 5998--6008.

\bibitem{dai2019transformer}
Zihang Dai, Zhilin Yang, Yiming Yang, Jaime Carbonell, Quoc~V Le, and Ruslan
  Salakhutdinov,
\newblock ``Transformer-xl: Attentive language models beyond a fixed-length
  context,''
\newblock in {\em ACL}, 2019, pp. 2978--2988.

\bibitem{panayotov2015librispeech}
Vassil Panayotov, Guoguo Chen, Daniel Povey, and Sanjeev Khudanpur,
\newblock ``Librispeech: an asr corpus based on public domain audio books,''
\newblock in {\em ICASSP}, 2015, pp. 5206--5210.

\bibitem{park2020specaugment}
Daniel~S Park, Yu~Zhang, Chung-Cheng Chiu, Youzheng Chen, Bo~Li, William Chan,
  Quoc~V Le, and Yonghui Wu,
\newblock ``Specaugment on large scale datasets,''
\newblock in {\em ICASSP}, 2020, pp. 6879--6883.

\bibitem{kudo2018sentencepiece}
Taku Kudo and John Richardson,
\newblock ``Sentencepiece: A simple and language independent subword tokenizer
  and detokenizer for neural text processing,''
\newblock in {\em EMNLP}, 2018, pp. 66--71.

\bibitem{pytorch}
Adam Paszke, Sam Gross, Francisco Massa, Adam Lerer, James Bradbury, Gregory
  Chanan, Trevor Killeen, Zeming Lin, Natalia Gimelshein, Luca Antiga, Alban
  Desmaison, Andreas Kopf, Edward Yang, Zachary DeVito, Martin Raison, Alykhan
  Tejani, Sasank Chilamkurthy, Benoit Steiner, Lu~Fang, Junjie Bai, and Soumith
  Chintala,
\newblock ``Pytorch: An imperative style, high-performance deep learning
  library,''
\newblock in {\em NeurIPS}, 2019, pp. 8024--8035.

\bibitem{kingma2014adam}
Diederik~P Kingma and Jimmy Ba,
\newblock ``Adam: A method for stochastic optimization,''
\newblock in {\em ICLR}, 2014.

\bibitem{jim1996analysis}
Kam-Chuen Jim, C~Lee Giles, and Bill~G Horne,
\newblock ``An analysis of noise in recurrent neural networks: convergence and
  generalization,''
\newblock {\em IEEE Transactions on neural networks}, vol. 7, no. 6, pp.
  1424--1438, 1996.

\bibitem{heafield2011kenlm}
Kenneth Heafield,
\newblock ``Kenlm: Faster and smaller language model queries,''
\newblock in {\em Proceedings of the sixth workshop on statistical machine
  translation}, 2011, pp. 187--197.

\end{thebibliography}

\clearpage

\appendix
\section{Additional Experiments}

\subsection{Model Scaling}
In order to study the effect of model scaling on recognition performance, we design larger Efficient Conformer CTC models of 31M and 125M parameters. Tables~\ref{table:HyperparametersEffConfCTC} describes architecture hyper-parameters of Efficient Conformer Small, Medium and Large CTC variants. We similarly identify Medium and Large Conformer CTC models within the same parameter range (Table~\ref{table:HyperparametersConformer}).
\begin{table}[ht]
    \centering
    \scriptsize
    \caption{Efficient Conformer CTC models hyper-parameters.}
    \begin{tabular}{c|ccc}
        \hline
        \multirow{2}{*}{Model} & Eff Conf & Eff Conf & Eff Conf\\
        & (S) & (M) & (L)\\\hline\hline
        Num Params (M)      & 13.2          & 31.5          & 125.6         \\
        Encoder Blocks      & 5,5,5         & 5,6,5         & 5,6,5         \\
        Encoder Dims        & 120,168,240   & 180,256,360   & 360,512,720   \\
        Attention Heads     & 4,4,4         & 4,4,4         & 8,8,8         \\
        Conv Kernel Size    & 15,15,15      & 15,15,15      & 15,15,15      \\
        Att Group Size      & 3,1,1         & 3,1,1         & 3,1,1         \\\hline
    \end{tabular}
    \label{table:HyperparametersEffConfCTC}
    \caption{Conformer CTC models hyper-parameters.}
    \begin{tabular}{c|ccc}
        \hline
        \multirow{2}{*}{Model} & Conformer & Conformer & Conformer\\
        & (S) & (M) & (L)\\\hline\hline
        Num Params (M)      & 13.0  & 30.5  & 121.5\\
        Encoder Blocks      & 16    & 18    & 18  \\
        Encoder Dim         & 176   & 256   & 512  \\
        Attention Heads     & 4     & 4     & 8    \\
        Conv Kernel Size    & 31    & 31    & 31    \\\hline
    \end{tabular}
    \label{table:HyperparametersConformer}
\end{table}

Table~\ref{table:results2} compares the Word Error Rates obtained on the LibriSpeech dataset with recently published CTC, Sequence-to-sequence (S2S) and Transducers approaches. Our Efficient Conformer CTC Large model trained on 4 Nvidia RTX 3090 GPUs achieves near state-of-the-art performance of 2.5\%/5.8\% without using a language model and 2.1\%/4.7\% with an external n-gram language model for test-clean/test-other. We find Small, Medium and Large Efficient Conformer CTC to reach lower word error rates than Citrinet models using an external 6-gram language model. However, this small gain in accuracy isn't sufficient to close the gap between SOTA Transducers approaches. We suppose that more computing resources and longer training should help to further reduce this gap and compare these approaches more equitably.   
\begin{table}[ht]
    \centering
    \scriptsize
    \caption{Comparison of LibriSpeech WER(\%) with recent published CTC, Seq2Seq and Transducer models.}
    \hfill \break
    \begin{tabular}{cccccc}
    \hline
    \multicolumn{1}{c}{\multirow{2}{*}{\begin{tabular}[c]{@{}c@{}}\textbf{Model} \\ \textbf{Architecture}\end{tabular}}} & \multicolumn{1}{c}{\multirow{2}{*}{\begin{tabular}[c]{@{}c@{}}\textbf{Model} \\ \textbf{Type}\end{tabular}}} & \multirow{2}{*}{\textbf{LM}} & \multicolumn{2}{c}{\textbf{test WER}} & \multirow{2}{*}{\begin{tabular}[c]{@{}c@{}}\textbf{Params}\\ \textbf{(M)}\end{tabular}} \\ \cline{4-5}
    \multicolumn{1}{c}{}                       &                           &  & \textbf{clean}       & \textbf{other}   &       \\ \hline
    Citrinet-256\cite{majumdar2021citrinet}    & CTC           & -         & 3.78  & 9.60   &       \\
                    &               & 6-gram    & 3.65  & 8.06  & 9.8   \\
                    &               & Trans-XL  & 2.75  & 6.87  &       \\ \hline
    Citrinet-384\cite{majumdar2021citrinet}    & CTC           & -         & 3.20  & 7.90  &       \\
                    &               & 6-gram    & 2.94  & 6.71  & 21.0  \\
                    &               & Trans-XL  & 2.52  & 5.95  &       \\ \hline
    Citrinet-512\cite{majumdar2021citrinet}     & CTC           & -                 & 3.11  & 7.82  &       \\
                                            &               & 6-gram            & 2.40  & 6.08  & 36.5  \\
                                            &               & Trans-XL    & 2.19  & 5.50  &       \\ \hline
                                            
Citrinet-768\cite{majumdar2021citrinet}     & CTC           & -                 & 2.57  & 6.35  &       \\
                                            &               & 6-gram            & 2.15  & 5.11  & 81  \\
                                            &               & Trans-XL          & 2.04  & 4.79  &       \\ \hline
                                            
Citrinet-1024\cite{majumdar2021citrinet}    & CTC           & -                 & 2.52  & 6.22  &       \\
                                           &               & 6-gram            & 2.10  & 5.06  & 142  \\
                                            &               & Trans-XL    & 2.00  & 4.69  &       \\\hline
LAS-6-1280\cite{park2020specaugment}        & S2S       & -                 & 2.6   & 6.0   & 360 \\
                                     &               & RNN               & 2.2   & 5.2   &   \\\hline
Conformer\cite{guo2021recent}               & CTC+S2S & Trans-XL               & 2.1   & 4.9   & 115 \\ \hline
Tranformer\cite{zhang2020transformer}       & Trans-T    & -                 & 2.4   & 5.6   & 139  \\
                                            &               & Trans       & 2.0   & 4.6   &       \\ \hline
ContextNet(S)\cite{han2020contextnet}       & RNN-T    & -                 & 2.9   & 7.0   & 10.8  \\
                                            &               & RNN               & 2.3   & 5.5   &       \\ \hline
                                            
ContextNet(M)\cite{han2020contextnet}       & RNN-T    & -                 & 2.4   & 5.4   & 31.4  \\
                                            &               & RNN               & 2.0   & 4.5   &       \\ \hline
                                            
ContextNet(L)\cite{han2020contextnet}       & RNN-T    & -                 & 2.1   & 4.6   & 112.7  \\
                                            &               & RNN               & 1.9   & 4.1   &       \\ \hline
Conformer(S)\cite{gulati2020conformer}      & RNN-T    & -                 & 2.7   & 6.3   & 10.3  \\
                                            &               & RNN               & 2.1   & 5.0   &       \\ \hline
                                            
Conformer(M)\cite{gulati2020conformer}      & RNN-T    & -                 & 2.3   & 5.0   & 30.7  \\
                                            &               & RNN               & 2.0   & 4.3   &       \\ \hline
                                            
Conformer(L)\cite{gulati2020conformer}      & RNN-T    & -                 & 2.1   & 4.3   & 118.8  \\
                                            &               & RNN               & 1.9   & 3.9   &       \\ \hline\hline
    Conformer(S) & CTC           & -         & 4.07  & 10.25 & 13.0\\
                    &               & 6-gram    & 2.88  & 7.25\\    \hline
    Conformer(M) & CTC           & -         & 3.27  & 8.42 & 30.5\\
                    &               & 6-gram    & 2.55  & 6.32\\    \hline
    Eff Conformer(S)   & CTC           & -         & 3.57  & 8.99  & 13.2\\ 
       &               & 6-gram    & 2.72  & 6.66\\\hline
    Eff Conformer(M)  & CTC    & -         & 2.96  & 7.57  & 31.5\\ 
       &                        & 6-gram    & 2.37  & 5.82\\\hline
    Eff Conformer(L)  & CTC    & -         & 2.54  & 5.79  & 125.6\\ 
       &                        & 6-gram    & 2.10  & 4.71\\\hline
    \end{tabular}
    \label{table:results2}
\end{table}

\subsection{Models Inference Time}

\begin{figure}[H]
    \centering
    \centerline{\includegraphics[width=\linewidth]{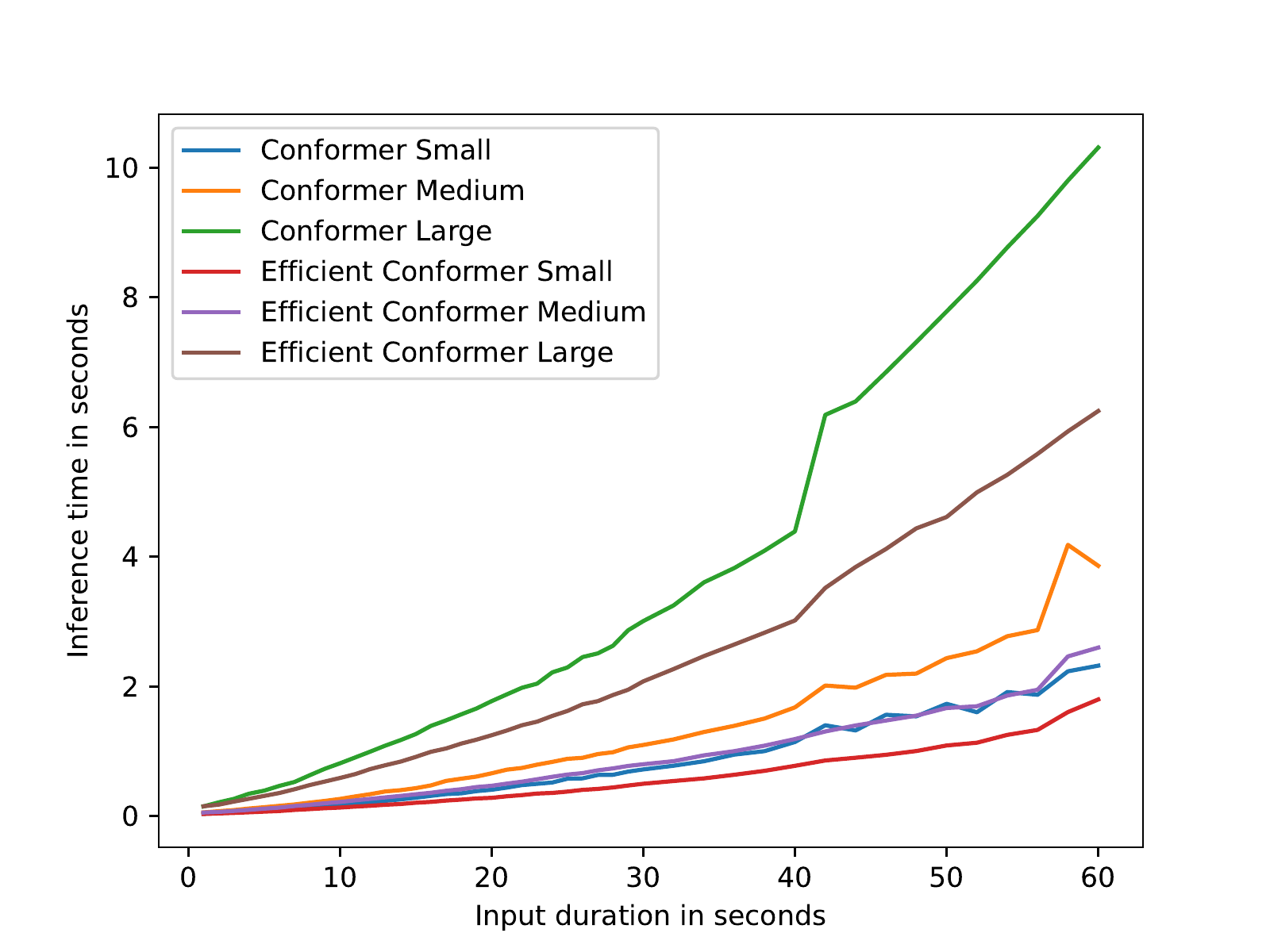}}
    \caption{CTC models inference time for processing long sequences measured on a single Intel Core i9-9940X 3.3GHz CPU thread.}
    \label{fig:TimesCPU}
\end{figure}

Figure~\ref{fig:TimesCPU} shows inference times of Conformer and Efficient Conformer CTC variants for different input lengths. The Efficient Conformer architecture greatly reduces inference time and allows us to reach better recognition performance while requiring less CPU time to process audio sequences. As shown is Table~\ref{table:results} and Table~\ref{table:results2}, we find Efficient Conformer CTC models to consistently outperform Conformer variants.

\subsection{Multi-Head Linear Self-Attention}

Linear attention, also known as efficient attention, has a linear memory and computational complexity with respect to the size of the input. It does not compute a similarity between each pair of positions but global context vectors for each feature dimension. This results in an $O(d^{2} \cdot n)$ computational complexity which bring an efficiency advantage over regular dot-product attention when $n$ is way larger than $d$. We study the use of multi-head linear self-attention as proposed in~\cite{li2021efficient} for our small progressively downsampled CTC model where feature dimension is relatively smaller than sequence length. Multi-head linear self-attention is defined as:
\begin{align}
MHLSA(X) = Concat\left(O_{1}, ..., O_{H}\right)W^{O},\\
\text{where}\ O_{h} = \sigma_{row}\left(\frac{Q_{h}}{{d_{h}}^{\frac{1}{4}}}\right)\left(\sigma_{col}\left(\frac{K_{h}}{{d_{h}}^{\frac{1}{4}}}\right)^{T}V_{h}\right)
\end{align}
Where $\sigma_{row}(\cdot)$ and $\sigma_{col}(\cdot)$ denote the operators of applying the softmax function along the rows and columns of a matrix. Table~\ref{table:LinearAttention} compares the use of linear attention in the progressively downsampled conformer encoder with regular dot-product attention. Linear attention greatly reduce inference and training time for our small model but also hurts recognition performance. A good trade off between accuracy and model complexity can be achieved using grouped attention in earlier layers. An interesting follow-up to this work would be to experiment using multi-head linear self-attention in earlier layers where hidden sequence length is relatively larger than model feature dimension and compare it with grouped attention.
\begin{table}[ht]
    \centering
    \scriptsize
    \caption{Ablation study on linear attention}
    \hfill \break
    \begin{tabular}{ccccccc}
        \hline
         \multicolumn{1}{c}{\multirow{2}{*}{\begin{tabular}[c]{@{}c@{}}\textbf{Attention} \\ \textbf{Type}\end{tabular}}} & \multicolumn{1}{c}{\multirow{2}{*}{\begin{tabular}[c]{@{}c@{}}\textbf{Model} \\ \textbf{Type}\end{tabular}}} & \multirow{2}{*}{\begin{tabular}[c]{@{}c@{}}\textbf{dev}\\ \textbf{clean}\end{tabular}} & \multirow{2}{*}{\begin{tabular}[c]{@{}c@{}}\textbf{dev}\\ \textbf{other}\end{tabular}} & \multirow{2}{*}{\begin{tabular}[c]{@{}c@{}}\textbf{MAdds} \\ \textbf{(B)}\end{tabular}} & \multirow{2}{*}{\begin{tabular}[c]{@{}c@{}}\textbf{Inv} \\ \textbf{RTF}\end{tabular}} & \multirow{2}{*}{\begin{tabular}[c]{@{}c@{}}\textbf{Train} \\ \textbf{Time (h)}\end{tabular}}\\\\ \hline
        Regular             & CTC   & 3.44  & 9.12  & 3.91  & 48.8  & 191\\
        Grouped (3,1,1)     & CTC   & 3.40  & 9.13  & 3.51  & 61.9  & 124   \\
        Grouped (9,5,3)     & CTC   & 3.56  & 9.74  & 3.16  & 68.9  & 113   \\
        Linear              & CTC   & 3.89  & 10.02 & 2.87  & 70.7  & 91    \\\hline
    \end{tabular}
    \label{table:LinearAttention}
\end{table}

\end{document}